\documentclass{osa-article}

\journal{osajournal}

\articletype{Research Article}

\newcommand{\revised}[1]		{{\color{black}#1}}

\begin{document}

\title{Simple and compact diode laser system stabilized to Doppler-broadened iodine-lines at 633 nm}

\author{F. Krause, \authormark{1,*}
  E. Benkler, \authormark{1}
	C. N\"olleke,	\authormark{2}
	P. Leisching, \authormark{2}
	U. Sterr  \authormark{1}}

\address{\authormark{1}Physikalisch-Technische Bundesanstalt, Bundesalle 100, 38116 Braunschweig, Germany \\
\authormark{2}TOPTICA Photonics AG, Lochhamer Schlag 19, 82166 Gr\"afelfing, Germany}

\email{\authormark{*}florian.krause@ptb.de}




\begin{abstract}
We present a compact iodine-stabilized laser system at 633~nm, based on a distributed-feedback laser diode. 
Within a footprint of $27\times 15$ cm$^2$ the system provides 5~mW of frequency stabilized light from a single-mode fiber. 
Its performance was evaluated in comparison to Cs clocks representing primary frequency standards, realizing the SI unit Hz via an optical frequency comb.  
With the best suited absorption line the laser reaches a fractional frequency instability below $10^{-10}$ for averaging times above 10~s. 
The performance was investigated at several iodine lines and a model was developed to describe the observed stability on the different lines.
\end{abstract}


\section{Introduction}

\revised{Due to their simplicity and reliability helium-neon (He-Ne) lasers at a wavelength of 633~nm are widely used} for interferometric length measurements\revised{ \cite{jae16} and metrology applications \cite{qui03}}.
Without any additional reference, Zeeman-stabilized and two-mode frequency stabilized He-Ne laser have shown instabilities of $2\cdot10^{-11}$ and $3\cdot10^{-10}$ respectively at averaging times of 1000~s, and frequency drifts of about $2 \cdot 10^{-8}$ over several months \cite{row90a,cid83} 
\revised{with typical output powers of less than a milliwatt.} 
These properties well meet the requirements of commercial laser interferometers.
He-Ne lasers with internal iodine cell stabilized to Doppler free molecular hyperfine lines of iodine achieve an instability down to $1 \cdot 10^{-13}$ at 1000~s averaging time, and an absolute uncertainty of $2.1 \cdot 10^{-11}$ \cite{BIPM18}.
The stabilization of these systems is more demanding and they are mostly used for calibration.    

However, He-Ne lasers require relatively large volumes even at low output powers, have a low power efficiency and they do not offer the possibility for wide-bandwidth frequency tuning.     
Furthermore, the technical know-how for building and maintaining He-Ne lasers is vanishing and hence alternative techniques in this wavelength range are needed.

Stabilizing a diode laser to a molecular or atomic reference is a promising substitute for He-Ne lasers as this eliminates their drawbacks \revised{\cite{gaw04}}.
A narrow linewidth 633~nm diode laser stabilized to Doppler broadened iodine absorption lines has reached an instability of $1\cdot 10^{-9}$ at 1000~s averaging time as evaluated with a wavelength meter. 
This laboratory setup employs an iodine cell with a length of 30~cm. 
Because of a low pressure (14~\textdegree C saturation temperature), a relatively long effective interaction length of 90~cm has to be used for the spectroscopy~\cite{rer17}.         
 
\revised{
Here we present a simple and compact shoe-box-sized diode laser system at 633~nm with fiber output, as direct one-to-one replacement of stabilized He-Ne lasers in industrial applications, such as interferometric length measuring systems or laser trackers \cite{mur16a}.
As these systems employ specific optical components, the operational wavelength of 633~nm is a mandatory requirement. 
This rules out the use of Diode-Pumped Solid State (DPSS) lasers or robust diode lasers stabilized to rubidium (Rb) at 780~nm.
So far available narrow-linewidth diode lasers at 633~nm were complex extended cavity diode lasers (ECDL) that are not robust enough for industrial application.  
In our system a robust distributed feedback (DFB) laser diode is stabilized to Doppler broadened iodine $\left( ^{127}\mathrm{I}_{2} \right)$ absorption lines and the system includes a double-stage isolator and a fiber coupling (Fig. \ref{fig:1}).
To achieve a compact design, Doppler free saturation spectroscopy is not used because a more complicated setup with larger cell and higher optical power to saturate the weak molecular transition would be required \cite{tal98}. 
This 633~nm diode laser system presented here with a relatively small iodine cell (3.3~cm length) reaches a fractional frequency instability of $1.9 \cdot 10^{-11}$ at 1000~s averaging time.} 
The stability of the laser system stabilized to different iodine lines was evaluated using an optical frequency comb.

\section{Experimental Setup}
\label{sec:examples}

\subsection{Stabilized laser system}

\begin{figure}[]
\centering
\includegraphics[width=0.7\linewidth]{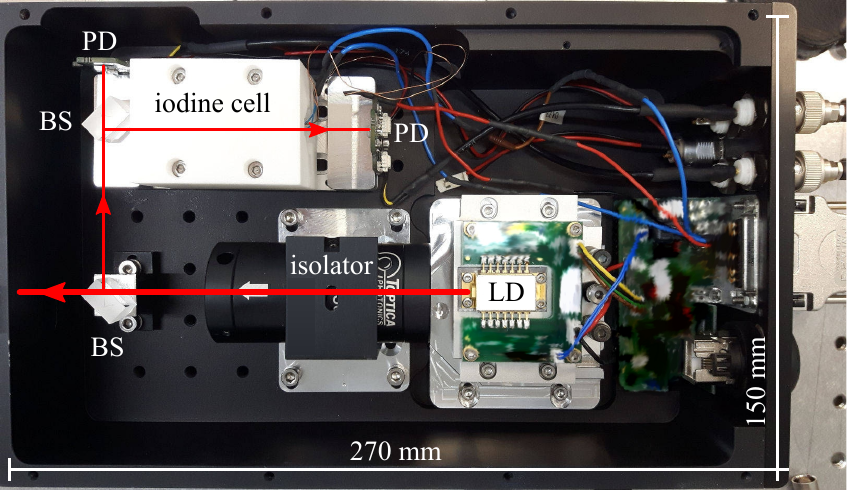}
\caption{Picture of the diode laser system with DFB laser diode (LD), isolator, beamsplitters (BS), photodetectors (PD) and iodine cell inside a temperature controlled environment. \revised{To the left a fiber coupler is attached to the housing (not shown here).}  }
\label{fig:1}
\end{figure}
\noindent
\begin{figure}[!h]
\centering
\includegraphics[width=\linewidth]{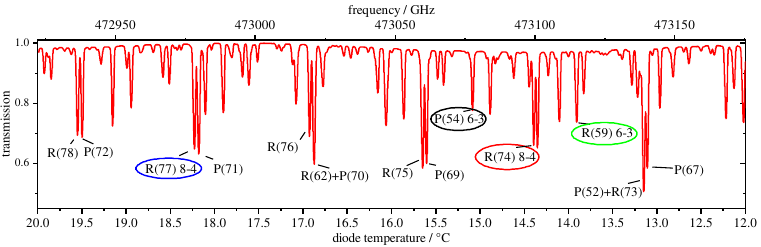}  
\caption{
Measured normalized iodine transmission spectrum as a function of the laser diode temperature. 
The iodine lines used for frequency stabilization are marked in color.
}
\label{fig:transmission}
\end{figure}

Fig. \ref{fig:1} shows the system in its 27~cm $\times$ 15~cm housing.
Light emitted from the DFB laser diode first passes an isolator to prevent perturbations by reflections back into the laser diode. 
Behind the isolator the light is split by a beamsplitter (BS) and
\revised{the main beam is coupled to a single-mode fiber by a fiber coupler mounted to the housing. 
A small part of the light is used for spectroscopy which is further split by a second beamsplitter.} 
The reflected part is sent through the iodine cell onto a photo diode (PD), while the transmitted beam is monitored by a second photo diode to provide a reference signal for normalization of the absorption spectrum. 
To achieve a strong absorption, the 3.3~cm long iodine cell (with purity according to the manufacturer >98~\%) was heated to a temperature of 60~\textdegree C.
\revised{This temperature is a good trade-off between strong absorption (of approximately 50~\%) for the strongest lines and small heating power.}
Using published iodine vapor pressure data \cite{hon60} interpolated by the Antoine equation \cite{tho46} we estimate a saturated iodine vapor pressure of 616~Pa inside the cell.

By varying the diode temperature, the optical frequency can be tuned over a range of $\Delta \nu=245$~GHz without mode hops to scan the iodine spectrum (Fig. \ref{fig:transmission}). 
\revised{The laser current has a much smaller impact on the laser frequency of about 1~GHz per 1~mA with significant power variation \cite{noe18a}. However, due to its high actuator bandwidth, the current is used in the lock to iodine to correct for fast laser frequency fluctuations.} 

Depending on the diode temperature, the power at the fiber output is between 4.5~mW and 6.5~mW. 
To stabilize the laser frequency to a peak of an absorption line via a 1f-lock-in technique, the laser current is modulated at a frequency $f_{\mathrm{mod}}=21$~kHz.
The corresponding peak-to-peak frequency modulation deviation of the emitted light was kept as small as $\Delta \nu_{\mathrm{mod}}=5$~MHz, which is much smaller than the Doppler-broadened iodine linewidth.

\revised{
When the laser is used for interferometry, this modulation leads to a phase modulation of the interference signal, depending on the path difference. 
Hence, the interference contrast is reduced if the data acquisition averages over the modulation. If we allow for a maximum peak-to-peak phase modulation of $\pi$, (contrast reduced to the $0^\mathrm{th}$ order Bessel function $J_{0}(\pi /2)\approx0.5$), this limits the path difference to 
$L_{\mathrm{coh,\pi}}\approx\frac{c}{2 \Delta \nu_{\mathrm{mod}}}=30\,\mathrm{m}$.  
Instead, if the data acquisition is fast enough to follow the modulation of the interference fringes, it will not limit the coherence length. 
The coherence length is then determined by the linewidth $\delta \nu$ of the laser diode, which is less than $<1.5\, \mathrm{MHz}$. 
Assuming a Lorentzian line profile, the coherence length is then
$L_{\mathrm{coh}}=\frac{c}{\pi \delta \nu } > 63\,\mathrm{m}$ \cite{sal91}. 
}

\revised{
The system automatically scans the iodine spectrum, identifies the correct absorption line and locks to that line within less than 5 minutes.
After initial power-on the system needs 10 to 15 minutes until all the parameters (especially the temperature of the iodine cell) are settled and the laser frequency is stabilized. 
}

\subsection{Frequency Measurement}

\begin{figure}[]
\centering
\includegraphics[width=0.7\linewidth]{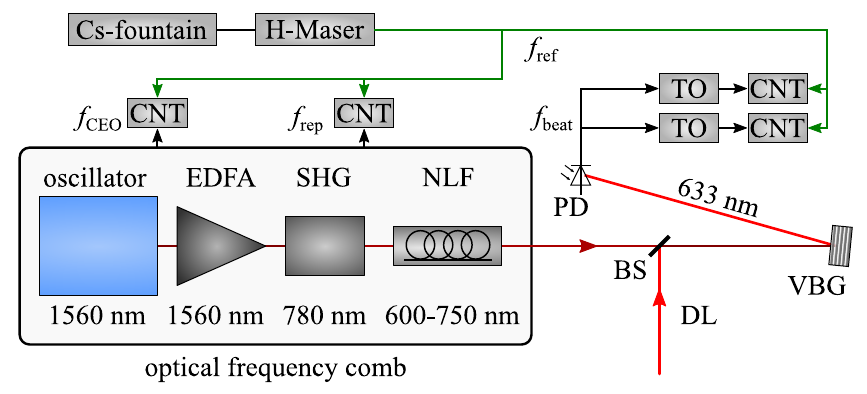}
\caption{\revised{
Sketch of the frequency comb and the experimental setup for the optical frequency measurement of the diode laser light (DL) with Er-doped fiber amplifier (EDFA), second harmonic generation (SHG), nonlinear fiber (NLF), beamsplitter (BS), volume Bragg grating (VBG), tracking oscillators (TO), frequency counters (CNT) and photo diode (PD).}
}
\label{fig:kamm}
\end{figure}
\begin{figure}[tbp]
\centering
\includegraphics[width=0.7\linewidth]{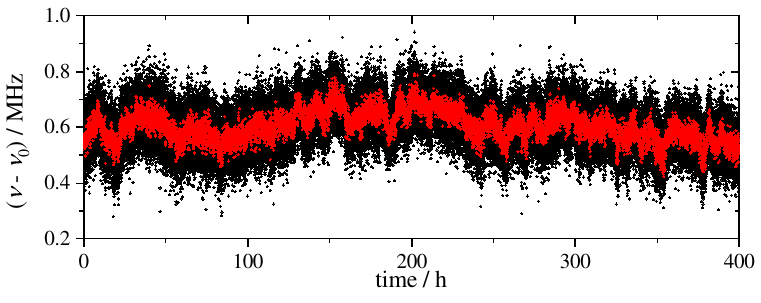}
\caption{Absolute frequency of the diode laser system stabilized to \revised{R(74) 8-4 } over several days averaged over 10 s (black) and 100 s (red),\revised{ with an offset of $\nu_0=473\,099\,403$~MHz subtracted.}}
\label{fig:beatmeasurement}
\end{figure}

\noindent
The long term instability and the absolute frequency $\nu$ of the laser system are characterized in comparison to two Cs fountain clocks via a hydrogen maser and an optical frequency comb (Fig.~\ref{fig:kamm}). 
The comb spectrum of the comb-generating Er:fiber-based fs-laser oscillator is centered around 1560~nm. 
After amplification in an Er-doped fiber amplifier (EDFA), the second harmonic comb spectrum around a wavelength of 780~nm is generated (SHG), which in a nonlinear fiber (NLF) generates a super-continuum spanning 600-750~nm \cite{hol00}. 
The fields of the super-continuum and the DFB-laser system (DL) are overlapped using a beamsplitter. 
With a volume Bragg grating (VBG) most of the comb lines besides those near the line of the DFB-laser at 633 nm are filtered out in order to improve the signal-to noise ratio of the beat note detected with a photo diode. 
For band-pass filtering of the radiofrequency beat signal, tracking oscillators~(TO) are phase-locked to the signal. Thus, clean signals are provided to the inputs of frequency counters (CNT) in $\Lambda$-averaging mode \cite{ben15} (K+K FXE) for dead-time-free synchronous measurement and recording of the RF frequencies. 
To make sure that the center frequency of the beat signal is tracked correctly despite the frequency modulation of the laser, different tracking oscillators with slightly different, asymmetrically chosen parameters and two counters are used. 
The frequency difference between these two channels is in the range of a few Hz ($\Delta\nu/\nu \approx 10^{-14}$), thus significant measurement errors due to cycle slips of the tracking filters can be excluded.
The optical absolute frequency $\nu$ is calculated from the beat frequency $f_{\mathrm{beat}}$:

\begin{equation}
\nu= 2 f_{\mathrm{CEO}}+n \cdot f_{\mathrm{rep}} + f_{\mathrm{beat}},
\label{eq:combfrequency}
\end{equation}
\noindent
where $f_{\mathrm{CEO}}$ is the carrier envelope offset frequency and $f_{\mathrm{rep}}$ the repetition rate.
Both frequencies are also recorded by K+K FXE counters.
\revised{
All counters (CNT) use a reference signal at frequency $f_{\mathrm{ref}}=10\,\mathrm{MHz}$ from an active hydrogen maser.
The H-maser is referenced to a Cs fountain clock, which is a primary frequency standard realizing the unit Hertz.
}
The mode number $n$ of the comb line and the sign of $f_\mathrm{beat}$ are determined from a rough frequency measurement with a wavelength-meter having a few ten MHz uncertainty.
Fig. \ref{fig:beatmeasurement} shows the measured absolute frequency of the laser stabilized to the line R(74) 8-4.

\subsection{Simulation of Doppler broadened iodine spectra}

Each absorption line in Fig. \ref{fig:transmission} consists of several hyperfine transitions with Voigt line shapes.
Their Gaussian widths are given by the Doppler broadening ($\delta \nu_{\mathrm{DB}} = 388\, \mathrm{MHz}$) and their Lorentzian widths contain natural and collision broadening. 
In the investigated frequency range, the upper state lifetime of iodine $^{127}\mathrm{I}_{2}$ is in the range of 300-400~ns \cite{sho72}, leading to a natural line width of about 400-500~kHz.  
At our vapor pressure and temperature collision broadening amounts to ($\delta \nu_{\mathrm{co}} = 76.6\, \mathrm{MHz}$) \cite{bri77}.
Compared to these broadening contributions, additional transit time broadening can be neglected.
For our simulations the Voigt profile was calculated as the real part of the Faddeeva function \cite{hui78}. 

The absorption lines around 633~nm are between rovibrational levels in the $X$ and $B$ potentials of iodine $^{127}\mathrm{I}_{2}$ and consist of several hyperfine lines in a range of about 1 GHz.
The number of hyperfine lines depends on the rotational quantum number $J''$ of the molecular ground state. Transitions with an even $J''$ have 15 hyperfine components and with an odd $J''$ 21, due to the required symmetry of the homonuclear $^{127}\mathrm{I}_{2}$ molecular wavefunction \cite{kro69}.

The frequency and the relative intensity of the iodine hyperfine transitions were calculated using the program IodineSpec \cite{bod02,kno04}.
This software is based on molecular potentials for the two electronic states involved and an interpolation for hyperfine splittings and achieves a standard (1-$\sigma$) frequency uncertainty of 1.5~MHz.
The transmission spectrum $T(\nu)$ of iodine in the frequency region of the laser system was simulated by summing the Voigt profiles of the individual hyperfine transitions. 
The relative intensities given by the program were scaled to match the simulated transmission to the experimental data (Fig.~\ref{fig:transmission}).

A pressure shift of about -5.9~MHz was included due to the high temperature of 333~K and corresponding vapor pressure of 616~Pa. The shift was obtained by scaling the pressure shift for a He-Ne laser stabilized to the R(127)~11-5 line of -9.5~kHz/Pa at a temperature of 288~K (18~Pa) \cite{edw99}. 

\begin{figure}[]
\centering
\includegraphics[width=0.7\linewidth]{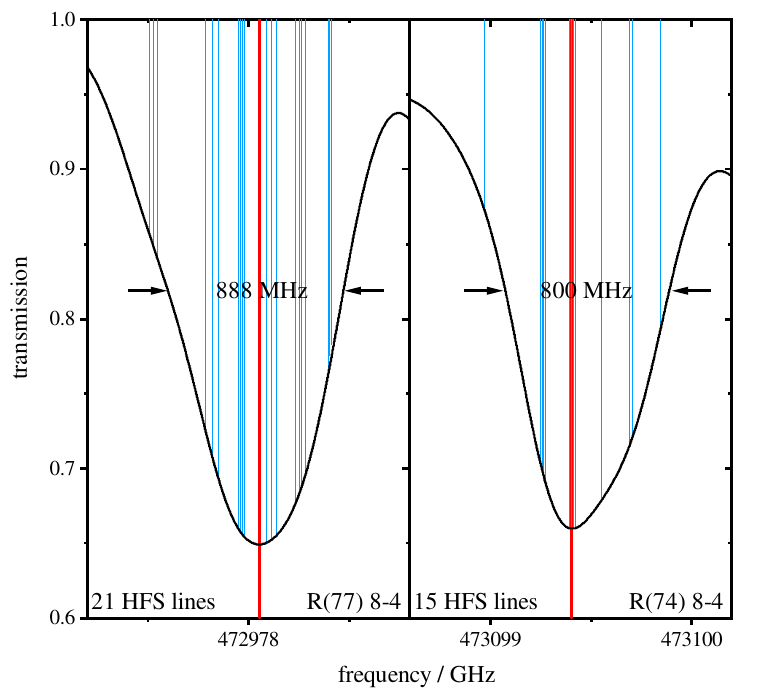}
\caption{Simulated transmission spectra of two Doppler-broadened iodine lines $^{127}\mathrm{I}_{2}$ with the frequencies of the HFS components (blue), the measured center frequency $\nu_{\mathrm{c}}$ of the laser system (red) and the FWHM. }
\label{fig:5}
\end{figure}

Fig. \ref{fig:5} presents the simulated line shapes of the transmission spectra of line R(77) 8-4 (21 HFS components) and R(74) 8-4 (15 HFS components). The two lines illustrate the influence of the number of HFS components on the total profile.
The profile with 15 HFS components is more sharply peaked and shows a smaller full width at half maximum (FWHM) compared to the one with 21 components. This behavior is typical for all lines in the tuning range of the diode laser.

\section{Experimental Results}

To investigate the frequency stabilized laser system, Doppler broadened lines with 15 hyperfine components (P(54)~6-3, R(74)~8-4) and lines with 21 components (R(59)~6-3, R(77)~8-4) were used as reference. 
\revised{For both kinds of lines we have chosen a strong line with a minimum transmission $T_0\approx 0.66$ and a weaker line with $T_0\approx 0.76$.}

\begin{table*}[]
\centering
\begin{tabular}{cccccccc}
\hline
Linie & HFS & $\nu_{\mathrm{c}}$ / MHz & $\nu_{\mathrm{sim}}$ / MHz  & $\Delta \nu$ / MHz & $\sigma_{y}(128\, \mathrm{s})$ & $T_{\mathrm{min}}$ / \% & $\kappa_{\mathrm{sim}}$ / $\mathrm{GHz}^{-2}$ \\
\hline
P(54) & 15 & 473\,077\,641.3(2)  & 473\,077\,638.4  &  2.9 & $4.7 \cdot 10^{-11}$  & 78 & 1.96\\
R(74) & 15 & 473\,099\,403.6(1)  & 473\,099\,406.6  & -3.0 & $3.0 \cdot 10^{-11}$  & 66 & 2.83 \\
R(59) & 21 & 473\,114\,765.8(4)  & 473\,114\,756.9  &  8.9 & $7.7 \cdot 10^{-11}$ & 74 & 1.08 \\
R(77) & 21 & 472\,978\,055.3(2)  & 472\,978\,054.5  &  0.8 & $5.5 \cdot 10^{-11}$ & 65 & 1.77 \\
\hline
\end{tabular}
\caption{Comparison of the number of hyperfine transitions (HFS), the measured center frequency of the laser $\nu_{\mathrm{c}}$, the frequency position of the simulated minimum $\nu_{\mathrm{sim}}$,  the difference $\Delta \nu=\nu_{\mathrm{c}}-\nu_{\mathrm{sim}}$, the modified Allan deviation $\sigma_{y}(\tau)$ at $\tau=128\, \mathrm{s}$, the measured minimum transmission $T_{\mathrm{min}}$ and the calculated curvature $\kappa_{\mathrm{sim}}$ for the four investigated iodine lines. }
  \label{tab:1}
\end{table*} 

Stabilized on line R(74) 8-4 the optical frequency was measured over several days.
For the other lines the measurements lasted about 2 hours.
For each of these iodine lines table \ref{tab:1} shows the number of HFS components, 
the measured center frequency $\nu_{c}$ calculated from the beat measurements data, 
the frequency position of the simulated minimum $\nu_{\mathrm{sim}}$ 
and the difference $\Delta \nu=(\nu_{\mathrm{c}}-\nu_{\mathrm{sim}})$ between simulation and experimental results.  
Compared to the line width of around 850~MHz, the residual frequency differences smaller than 10~MHz represent a good agreement.

\begin{figure}[]
\centering
\includegraphics[width=0.7\linewidth]{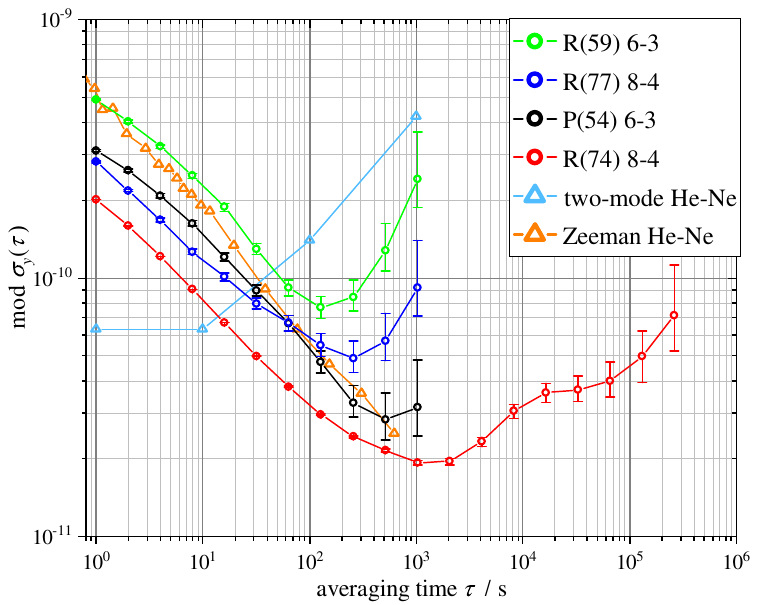}
\caption{Modified Allan deviation mod $\sigma_{y}(\tau)$ of the laser system stabilized to four different Doppler-broadened iodine lines. The data for the two-mode He-Ne are taken from~\cite{cid83} and for the Zeeman-stabilized He-Ne from~\cite{row90a}.}
\label{fig:4}
\end{figure}

Fig. \ref{fig:4} shows the modified Allan deviation mod $\sigma_{y}(\tau)$ of the measured diode laser frequency as a function of the averaging time $\tau$ for the four investigated lines. 
At short averaging times the instability decreases proportionally to $1/\sqrt{\tau}$ as expected for white frequency noise. 
However, at longer averaging times $\tau > 100 \ldots 1000$~s it starts to rise again. 

The short term instability can be explained by the signal-to-noise ratio of the error signal.
\revised{To generate the 1f-error signal, the laser frequency is modulated with rms amplitude $\Delta \nu_{ \mathrm{mod}}^{\mathrm{rms}}$
near the tip of the peak. In the neighborhood of the transmission minimum $T_0$ at frequency  $\nu_0$, the transmission can be approximated as 
$T(\nu) = T_0 + \kappa_\mathrm{sim} / 2 (\nu-\nu_0)^2$. 
The error signal is given as the 1f-rms component of the corresponding photocurrent $I_s$:}
\begin{equation}
 I_s(\nu) =     \frac{\eta P_{0} e}{h \nu} \kappa_\mathrm{sim} (\nu-\nu_0)  \Delta\nu_{\mathrm{mod}}^{\mathrm{rms}}.
\end{equation}

 Thus the slope $D = dI_s/d\nu$ of the error signal is proportional to the curvature $\kappa_\mathrm{sim}$ at the peak:

\begin{equation}
D =    \kappa_\mathrm{sim}   \frac{\eta P_{0}  e}{h \nu}  \Delta\nu_{\mathrm{mod}}^{\mathrm{rms}}.
\end{equation}

\revised{Here $P_{0}$ denotes the power at the input of the iodine cell and $\eta$ the detector quantum efficiency.
Using the slope $D$ and the noise spectral power density $S_I$ of the photocurrent, the short term instability for a laser frequency $\nu_{\mathrm{L}}$ can be estimated \cite{aff05} as:}

\begin{equation}
\sigma_{y}(\tau)=\frac{\sqrt{S_I}}{\sqrt{2}\, D \, \nu_{\mathrm{L}}}\tau^{-1/2}.
\label{eq:instabilitylimit}
\end{equation}
   
\revised{The stability is thus inversely proportional to the transmission curvature of the absorption peak.}   
The width close to the peak is smaller for 15 HFS lines compared to the peak with 21 lines and therefore their curvature is increased by a factor of two. 
For all investigated lines the curvature $\kappa_{\mathrm{sim}}$ is calculated from the simulated line shapes.
Line R(59) has the smallest peak curvature with $\kappa_{\mathrm{sim}}=1.09\, \mathrm{GHz^{-2}}$ followed by line R(77) ($\kappa_{\mathrm{sim}}=1.77\, \mathrm{GHz^{-2}}$), P(54) ($\kappa_{\mathrm{sim}}=1.96\, \mathrm{GHz^{-2}}$) and R(74) ($\kappa_{\mathrm{sim}}=2.83\, \mathrm{GHz^{-2}}$). 
This order is also visible in the modified Allan deviation at intermediate averaging times (Fig. \ref{fig:4}).
\revised{
The measured short term instability of the line R(74) ($\sigma_{y}(1\,\mathrm{s})=2\cdot 10^{-10}$) can be compared with the fundamental limit due to photon shot noise of the detected light. 
With an incident power of $P_{0}T_{0}=58\, \mathrm{\mu W}$ at the photodetector, the photocurrent shot noise amounts to 
$S_I = \frac{2 e^2 \eta T_{0} P_{0}  }{h \nu}$, 
resulting in the shot noise limited instability of 
$\sigma_{y}(\tau) = 2.6\cdot 10^{-11}\, (\tau / \mathrm{s})^{-1/2}$. 
This instability is a magnitude smaller than the measured Allan deviation at 1~s. 
Measuring the relative intensity noise spectrum of the laser diode, we discovered that in the frequency range near the modulation frequency, electronic noise of the detector is about a factor of ten above the shot noise. 
With this noise an instability of $\sigma_{y}(\tau)=1.4\cdot 10^{-10}\, (\tau / \mathrm{s})^{-1/2}$ would be reached,
which is in good agreement with the measured short term instability.} 

\revised{
The fact that the observed frequency instability is determined by the lock to iodine is further supported by analyzing the free running laser fractional frequency noise. 
We observe that the power spectral density of the free running laser frequency fluctuations $S_y$ for small frequencies 
($f<10\,\mathrm{kHz}$) shows flicker noise behavior ($S_{y}(f) =  4.5\cdot 10^{-19} f^{-1}$). 
This leads to a constant modified Allan deviation for the unstabilized laser 
$\sigma_{y} (\tau) = 6.5\cdot 10^{-10}$ \cite{daw07} 
at averaging times $\tau$ below 1 s.
The measured short term instability of the diode laser stabilized to line R(74) is below this value, which indicates that the short term stability is limited by the stabilization to the iodine vapor cell.}

Much stronger variations between the lines are seen in the long term stability.
Stabilized to line R(74)~8-4 (red) that shows the highest SNR, the laser system achieves the best frequency stability of all compared lines with a modified Allan deviation of $\sigma_{y}=2.0 \cdot 10^{-10}$ at an averaging time $\tau=1 \, \mathrm{s}$ and $\sigma_{y}=1.9 \cdot 10^{-11}$ at $\tau=1000 \, \mathrm{s}$.     
When stabilized to the second line with 15 HFS components, the Allan deviation rises at $\tau=(600\,\mathrm{s}$ - $1500\, \mathrm{s})$, while 
on the lines with 21 HFS it rises already at $\tau=(100\,\mathrm{s}$ - $200\, \mathrm{s})$.
We attribute this behavior to the different susceptibility of these lines to environmental perturbations.

\section{Conclusion}
 
We have presented a compact, iodine stabilized diode laser system at 633 nm with relative frequency instability below $10^{-10}$ ($2 \cdot 10^{-11}$ at $\tau=1000\, \mathrm{s}$), which is competitive to Zeeman- and two-mode stabilized He-Ne lasers.
In addition, we have shown that the laser can be tuned over a wide frequency range so that a large number of possible Doppler-broadened iodine lines can be used.
The absolute frequency and the observed behavior of the stability was in good agreement with simulations based on molecular potentials of iodine.

We found a significant dependence of the stability on the hyperfine structure of the Doppler-broadened absorption lines that were used for stabilization.
Because of its small size, lower electrical power consumption and high optical output power such stabilized diode laser systems using an external iodine-cell can become a valuable alternative to Zeeman- or two-mode stabilized He-Ne lasers at 633 nm.

\section{Funding Information}


This work is supported by the BMBF in the framework of the program KMU-innovativ: Photonik "FinDLiNG" (FKZ 13N13954) and by the EMPIR project 17FUN03 "USOQS". 
EMPIR projects are co-funded by the European Union's Horizon 2020 research and innovation programme and the EMPIR Participating States.
	
\section{Disclosures}
CN, PL: TOPTICA Photonics (E).


\begin{thebibliography}{10}
\newcommand{\enquote}[1]{``#1''}

\bibitem{jae16}
G.~J{\"a}ger, E.~Manske, T.~Hausotte, A.~M{\"u}ller, and F.~Balzer,
  \enquote{Nanopositioning and nanomeasuring machine {NPMM}-200{\textemdash}a
  new powerful tool for large-range micro- and nanotechnology,}
  {\protect\JournalTitle{Surface Topography: Metrology and Properties}}
  \textbf{4}, 034004 (2016).

\bibitem{qui03}
T.~J. Quinn, \enquote{Practical realization of the definition of the metre,
  including recommended radiations of other optical frequency standards
  (2001),} {\protect\JournalTitle{Metrologia}} \textbf{40}, 103--133 (2003).

\bibitem{row90a}
W.~R.~C. Rowley, \enquote{The performance of a longitudinal {Z}eeman-stabilised
  {He-Ne} laser (633 nm) with thermal modulation and control,}
  {\protect\JournalTitle{Measurement Science and Technology}} \textbf{1},
  348--351 (1990).

\bibitem{cid83}
P.~E. Ciddor and R.~M. Duffy, \enquote{Two-mode frequency-stabilised {He-Ne}
  (633 nm) lasers: studies of short- and long-term stability,}
  {\protect\JournalTitle{Journal of Physics E: Scientific Instruments}}
  \textbf{16}, 1223--1227 (1983).

\bibitem{BIPM18}
BIPM, \enquote{Recommended values of standard frequencies,} online at
  https://www.bipm.org/en/publications/mises-en-pratique/standard-frequencies.html
  (Page last updated: 30 November 2018).

\bibitem{gaw04}
W.~Gawlik and J.~Zachorowski, \enquote{Stabilization of diode-laser frequency
  to atomic transitions,} {\protect\JournalTitle{Optica Applicata}}
  \textbf{34}, 607--618 (2004).

\bibitem{rer17}
S.~Rerucha, A.~Yacoot, T.~M. Pham, M.~Cizek, V.~Hucl, J.~Lazar, and O.~Cip,
  \enquote{Laser source for dimensional metrology: investigation of an iodine
  stabilized system based on narrow linewidth 633 nm {DBR} diode,}
  {\protect\JournalTitle{Meas. Sci. Technol.}} \textbf{28}, 045204 (2017).

\bibitem{mur16a}
B.~Muralikrishnan, S.~Phillips, and D.~Sawyer, \enquote{Laser trackers for
  large-scale dimensional metrology: A review,}
  {\protect\JournalTitle{Precision Engineering}} \textbf{44}, 13 -- 28 (2016).

\bibitem{tal98}
H.~Talvitie, M.~Merimaa, and E.~Ikonen, \enquote{Frequency stabilization of a
  diode laser to {D}oppler-free spectrum of molecular iodine at 633 nm,}
  {\protect\JournalTitle{Opt. Commun.}} \textbf{152}, 182--188 (1998).

\bibitem{hon60}
R.~Honig and H.~Hook, \enquote{Vapor pressure data for some common gases,}
  {\protect\JournalTitle{RCA Rev.}} \textbf{21}, 360--368 (1960).

\bibitem{tho46}
G.~W. Thomson, \enquote{The {A}ntoine equation for vapor-pressure data.}
  {\protect\JournalTitle{Chemical Reviews}} \textbf{38}, 1--39 (1946).

\bibitem{noe18a}
C.~N{\"o}lleke, P.~Leisching, G.~Blume, D.~Jedrzejczyk, J.~Pohl, D.~Feise,
  A.~Sahm, and K.~Paschke, \enquote{Frequency locking of compact laser-diode
  modules at 633 nm,} in \emph{Photonic Instrumentation Engineering V,}  vol.
  10539 Y.~G. Soskind, ed., International Society for Optics and Photonics
  (SPIE, 2018), pp. 28--33.

\bibitem{sal91}
B.~E.~A. Saleh and M.~C. Teich, \emph{Fundamentals of Photonics} (John Wiley \&
  Sons, Inc., New York, New York, 1991).

\bibitem{hol00}
R.~Holzwarth, T.~Udem, T.~W. H\"ansch, J.~C. Knight, W.~J. Wadsworth, and
  P.~S.~J. Russell, \enquote{Optical frequency synthesizer for precision
  spectroscopy,} {\protect\JournalTitle{Phys. Rev. Lett.}} \textbf{85},
  2264--2267 (2000).

\bibitem{ben15}
E.~Benkler, C.~Lisdat, and U.~Sterr, \enquote{On the relation between
  uncertainties of weighted frequency averages and the various types of allan
  deviations,} {\protect\JournalTitle{Metrologia}} \textbf{52}, 565 (2015).

\bibitem{sho72}
K.~C. Shotton and G.~D. Chapman, \enquote{Lifetimes of $^{127}${I}$_2$
  molecules excited by the 632.8 nm {He/Ne} laser,} {\protect\JournalTitle{J.
  Chem. Phys.}} \textbf{56}, 1012--1013 (1972).

\bibitem{bri77}
A.~Brillet and P.~Cerez, \enquote{Quantitative description of the saturated
  absorption signal in iodine stabilized {He}-{Ne} lasers,}
  {\protect\JournalTitle{Metrologia}} \textbf{13}, 137 (1977).

\bibitem{hui78}
A.~K. Hui, B.~H. Armstrong, and A.~A. Wray, \enquote{Rapid computation of the
  {V}oigt and complex error functions,} {\protect\JournalTitle{J. Quant.
  Spectrosc. Radiat. Transfer}} \textbf{19}, 509 -- 516 (1978).

\bibitem{kro69}
M.~Kroll, \enquote{Hyperfine structure in the visible molecular-iodine
  absorption spectrum,} {\protect\JournalTitle{Phys. Rev. Lett.}} \textbf{23},
  631--633 (1969).

\bibitem{bod02}
B.~Bodermann, H.~Kn\"ockel, and E.~Tiemann, \enquote{Widely usable
  interpolation formulae for hyperfine splittings in the $^{127}${I}$_2$
  spectrum,} {\protect\JournalTitle{Eur. Phys. J. D}} \textbf{19}, 31--44
  (2002).

\bibitem{kno04}
H.~Kn{\"o}ckel, B.~Bodermann, and E.~Tiemann, \enquote{High precision
  description of the rovibronic structure of the {I}$_2$ {B}--{X} spectrum,}
  {\protect\JournalTitle{Eur. Phys. J. D}} \textbf{28}, 199--209 (2004).

\bibitem{edw99}
C.~S. Edwards, G.~P. Barwood, P.~Gill, and W.~R.~C. Rowley, \enquote{A 633 nm
  iodine-stabilized diode-laser frequency standard,}
  {\protect\JournalTitle{Metrologia}} \textbf{36}, 41 (1999).

\bibitem{aff05}
C.~Affolderbach and G.~Mileti, \enquote{A compact laser head with
  high-frequency stability for {Rb} atomic clocks and optical instrumentation,}
  {\protect\JournalTitle{Rev. Sci. Instrum.}} \textbf{76}, 073108 (2005).

\bibitem{daw07}
S.~T. Dawkins, J.~J. McFerran, and A.~N. Luiten, \enquote{Considerations on the
  measurement of the stability of oscillators with frequency counters,}
  {\protect\JournalTitle{IEEE Trans. Ultrason. Ferroelectr. Freq. Control}}
  \textbf{54}, 918--925 (2007). 

\end{thebibliography}



\end{document}